\def\PR{ Phys. Rev. }
\def\PRB#1{ Phys. Rev. {\bf #1} }
\def\JPCM#1{J. Phys. : Condens. Matter {\bf #1},}
\def\PRL#1{ Phys. Rev. Lett. {\bf #1}}

\documentclass[twocolumn,showpacs,preprintnumbers,amsmath,amssymb]{revtex4}
\usepackage{graphicx}
\usepackage{dcolumn}

  \def\ket{\vert \vert  \{ \emptyset \} \rangle}
  \def\ket2{\vert \vert \otimes \{ R \} \rangle}

  \def\ket{\vert \vert  \{ \emptyset \} \rangle}
  \def\ket2{\vert \vert \otimes \{ R \} \rangle}

\def\.#1{\mathaccent 95#1}
\def\^#1{\mathaccent 94 #1}
\def\~#1{\mathaccent "7E #1}

\def\pls{\enskip +\enskip}

  \def\ket{\vert \vert  \{ \emptyset \} \rangle}
  \def\ket2{\vert \vert \otimes \{ R \} \rangle}

\def\.#1{\mathaccent 95#1}
\def\^#1{\mathaccent 94 #1}
\def\~#1{\mathaccent "7E #1}

\def\pls{\enskip +\enskip}

  \def\ket{\vert \vert  \{ \emptyset \} \rangle}
  \def\ket2{\vert \vert \otimes \{ R \} \rangle}
  
\def\k{{\bf k}}

\begin{document}

\title {An augmented space approach to the study of phonons in disordered alloys : comparison between the itinerant coherent-potential approximation and the augmented space recursion}
\author{Aftab Alam}
\affiliation{S.N. Bose National Center for Basic Sciences, JD Block, Sector III, Salt Lake City, Kolkata 700 098, India }
\author{Subhradip Ghosh}
\affiliation{Department of Physics, Indian Institute of Technology, Guwahati, Assam, India}
\author{Abhijit Mookerjee}
\affiliation{S.N. Bose National Center for Basic Sciences, JD Block, Sector III, Salt Lake City, Kolkata 700 098, India }
\date{\today}
\begin{abstract}
A first principles density functional based linear response theory (the so called Density Functional Perturbation theory \cite{dfpt}) has been combined separately with two recently developed formalism for a systematic study of the lattice dynamics in disordered binary alloys. The two formalisms are the Augmented space recursion (ASR) \cite{alam} and the Itinerant coherent potential approximation (ICPA)\cite{glc}. The two different theories (DFPT-ASR and DFPT-ICPA) systematically provides a hierarchy of improvements upon the earlier single site based theories (like CPA etc.) and includes non-local correlations in the disorder configurations. The formalisms explicitly take into account fluctuations in masses, force constants and scattering lengths. The combination of DFPT with these formulation helps in understanding the actual interplay of force constants in alloys. We illustrate the methods by applying to a fcc Fe$_{50}$Pd$_{50}$ alloy. 
\end{abstract}
\maketitle
\section{ Introduction}
The last thirty years have seen  numerous attempts at setting up
a quantitatively accurate theory of phonons in disordered alloys. One of the earliest successful approximations was the coherent potential approximation \cite{taylor} (CPA). This approximation was a considerable improvement on the existing theories and, in examples of homogeneous disorder, was shown to yield configuration averaged Green functions which maintained lattice translational symmetry and  the herglotz analytical properties \cite{her} essential for physical interpretation \cite{mulhart}.
Despite its success, particularly in the electronic problem, the CPA was a single-site, mean-field approximation and could deal with only   diagonal (or mass, in the case of phonons)  disorder.
The phonon problem is  specifically difficult because, in it, diagonal 
 and off-diagonal disorders  are
impossible to separate. Moreover, the sum rule satisfied between the diagonal and off-diagonal parts of the force-constants  leads to environmental
disorder. That is, a
configuration fluctuation at a site affects the diagonal part of the dynamical matrix at its neighbours. Consequently, we do not expect the CPA
to give an adequate description of the phonon problem. That this was 
so  was discussed 
 in several papers \cite{kuni}-\cite{smith},  which indicated large
discrepancies between the CPA predictions and experimental results.  In the electron problem too, whenever there was off-diagonal disorder, as in the case of alloys with large size difference between its constituents  
leading to local lattice distortions \cite{latdis} or environmental disorder as in the case of alloys with short-range order \cite{mp}-\cite{durga}, the CPA was found to be inadequate. 

The hunt for adequate extensions of the
CPA was quite rigorous during the seventies and eighties \cite{shiba}-\cite{gonis}. 
Most of these generalizations were valid for very special types of off-diagonal disorder, which were mostly unphysical, or violated translational symmetry and herglotz properties.
Eventually, three approaches emerged as the most successful. Two of them were
 based on the augmented space theorem
of Mookerjee \cite{ast} : the itinerant
coherent-potential approximation (ICPA) of Ghosh {\em {et al}} \cite{glc} and
the augmented space recursion (ASR) of Saha {\em {et al}} \cite{sdm1} and Alam {\em {et al}} \cite{alam}. The former was an extension of the ideas of Mills and Ratanavaraksa \cite{mr} and
Kaplan {\em {et al}} \cite{klgd} and the latter combined the augmented space technique with the recursion method of Haydock {\em {et al}} \cite{hhk}. 
The third was a  very different and rather striking
approach  developed by Rowlands {\em {et al}} \cite{row2} and Biava {\em {et al}} \cite{biava} (the non-local CPA or NL-CPA) using the idea of {\sl coarse graining}
in reciprocal space originally proposed by Jarell and Krishnamurthy \cite{jar}.

More importantly a first principles ab-initio theory of phonons in disordered alloys is still lacking. Such a theory is needed in order to gain a microscopic understanding of the interplay of force constants in the complex phenomenon of phonon excitations.
Our aim in this communication is twofold : First, we shall discuss the similarities and differences between
the two methods (ICPA and ASR) based on the augmented space theorem. We shall apply both the techniques to identical models of an alloy system, FePd,  and discuss the comparison between their results. 
 Secondly, we shall  estimate the dynamical matrices from a first-principles approach to the parent ordered alloys and compare the ICPA and ASR results with experiment. We shall argue that first-principles estimates of the 
dynamical matrices on ordered versions do not yield  quantitativly accurate results (in comparison with experiment) for the disordered alloys. We shall propose that we need to go beyond and estimate the dynamical matrices from a model of embedded atoms in a fully disordered background. 

The outline of this paper is as follows. In section (II), we shall discuss in brief what the augmented space theorem is and its application to the problem of phonon excitations in disordered alloys. In section (III), we shall briefly describe the salient features of Itinerant coherent potential approximation (ICPA) and the Augmented space recursion (ASR). Section (IV) shall be devoted for the details of first principles calculation of force constants in alloys. The numerical results shall be discussed in section (V). Finally conclusions are drawn in section (VI)

\section{The Augmented Space Theorem}

The augmented space theorem has been discussed in great detail elsewhere \cite{TF}.
In this section we shall introduce only those  salient features 
 which will be required by us to understand the mathematics and notations in our
subsequent discussions.

For a homogeneously disordered binary alloy  the
Green function may be written as :

\begin{widetext}
\begin{eqnarray}
\ll \mathbf{G}(\mathbf{k},\omega^2)\gg & = & \frac{1}{N}\sum_{R,R'} \exp\left\{i\rule{0mm}{3mm}\mathbf{k}\cdot (R-R')\right\}\ \ll\rule{0mm}{4mm} \langle R\vert (\mathbf{m} \omega^2-\mathbf{\Phi})^{-1}\vert R'\rangle\gg,\nonumber\\
\mathbf{m} & = &\left\{\rule{0mm}{3mm} \mathbf{m}^A\ n_R + \mathbf{m}^B\ (1-n_R)\right\}\ \delta_{RR'}, \nonumber\\
\mathbf{\Phi}  & = &\mathbf{\Phi}^{AA}_{R-R'}\ n_R\ n_{R'}+ \mathbf{\Phi}^{BB}_{R-R'} \ (1-n_R)(1-n_{R'})+ \mathbf{\Phi}^{AB}_{R-R'}\ \left\{ \rule{0mm}{3mm} n_R(1-n_{R'})+(1-n_R)n_{R'}\right\},\nonumber\\
\end{eqnarray}
\end{widetext}

Here, $R,R'$ refer to lattice positions, $\ll\quad\gg$ refers to configuration averaging over random variables in the problem.  We should note here that the
Fourier transform in the first equation may be taken {\sl only} after 
the configuration averaging is carried out. The mass and force-constant matrices are matrices in the mode space and for systems with one atom per unit cell they are 3$\times$3.
$\{n_R\}$ are the random site-occupation variables which take values 1 and 0 
 depending upon whether the site labelled by $R$ is occupied by $A$ or
$B$-type of atom. The atom sitting at $R$ can either be of type $A \ (n_R=1)$
with probability $x$ or $B \ (n_R=0)$ with probability $y$.
The augmented space
formalism (ASF) now introduces the space of configurations of the set of binary 
random variables  $\{n_R\}$ : $\Psi$. 

In the absence of short-ranged order, each random variable  $n_R$ has associated with it an operator  ${\bf M}_R$
whose spectral density is its probability density :

\begin{eqnarray}
p(n_R) &=& x \delta(n_R-1) + y\delta(n_R) \nonumber \\ 
&=& -\frac{1}{\pi} \lim_{\delta\rightarrow 0} \ \mbox{Im} 
\langle \uparrow_R|\left((n_R+i\delta) {\bf I}-{\bf M}_R\right)^{-1}|\uparrow_R\rangle, \nonumber\\ 
\end{eqnarray}

\noindent where $x$,$y$ are concentrations of the components A and B, ${\bf M}_R$ is an operator whose eigenvalues $1, 0$ correspond to the observed values of $n_R$ and
whose corresponding eigenvectors $\{|1_R\rangle, |0_R\rangle\}$ span a configuration space
$\mathbf{\psi}_R$ of rank 2.  We may change the basis to  $\{|\uparrow_R\rangle,|\downarrow_R\rangle\}$ 
\begin{eqnarray*}
 \vert\uparrow_R\rangle = \left\{ \sqrt{x}\vert 1_R\rangle+\sqrt{y}\vert 0_R\rangle\right\}, \\
 \vert\downarrow_R\rangle = \left\{ \sqrt{y}\vert 1_R\rangle-\sqrt{x}\vert 0_R\rangle\right\}. 
\end{eqnarray*}
and in the new basis the operator ${\bf M}_R$ corresponding to $n_R$ is~: 
\begin{eqnarray*}
 {\bf M}_R &= & x{\cal P}^\uparrow_R + y{\cal P}^\downarrow_R + \sqrt{xy} \ {\cal T}^{\uparrow\downarrow}_R,\\
\end{eqnarray*}
\noindent  where ${\cal P}^{\uparrow}_R = \vert {\uparrow}_R\rangle\langle {\uparrow}_R\vert$, 
 ${\cal P}^{\downarrow}_R = \vert {\downarrow}_R\rangle\langle {\downarrow}_R\vert$  and 
 ${\cal T}^{\uparrow\downarrow}_R = \vert {\uparrow}_R\rangle\langle {\downarrow}_R\vert +   
 \vert {\downarrow}_R\rangle\langle {\uparrow}_R\vert$ are the projection and transfer operators in the configuration space $\mathbf{\psi}_R$ spanned by the two basis vectors. 
\vskip 0.2cm
 The full configuration space $\mathbf{\Psi}$ = $\prod^\otimes_R\ \mathbf{\psi}_R$ is then spanned by vectors of the form $\vert\uparrow\uparrow\downarrow\uparrow\downarrow\ldots\rangle$.
These configurations may be labelled by the sequence of sites $\{{\cal C}\}$ 
 at which we have a $\downarrow$. For example, for the state just quoted  $\{{\cal C}\}$
= $\vert\{3,5,\ldots\}\rangle$. This sequence is called the {\sl cardinality sequence}.
If we define the configuration $\vert\uparrow\uparrow\ldots\uparrow\ldots\rangle$ as the 
 {\sl reference} configuration, then the {\sl cardinality sequence} of the {\sl reference} configuration is the null sequence 
$\{\emptyset\}$.

In the full augmented space the operator corresponding to $n_R$ is :

\[
\widehat{\bf M}_R = {\cal I}\otimes \ldots \otimes{\mathbf M}_R\otimes\ldots \otimes{\cal I}\otimes\ldots\ \in\  \mathbf{\Psi}. \]

The augmented space theorem \cite{ast} then yields :
\begin{equation}
\ll {\bf G}(\k,\omega^2)\gg \ = \ \langle \k\otimes\{\emptyset\} |{(\widehat{\mathbf{m}}\omega^2 
- \widehat{\mathbf{\Phi}})}^{-1} |\k\otimes\{\emptyset\} \rangle,
\label{eq6}
\end{equation}
\noindent where the augmented {\bf k-} space basis $|\k\otimes\{\emptyset\}\ \rangle$ has the form

\[ (1/\sqrt{N})\sum_R \mbox{exp}(-i\k\cdot R)|R\otimes\{\emptyset\}\rangle. \]
The augmented space operators $\widehat{\mathbf m}$ and $\widehat{\mathbf{\Phi}}$ are constructed from the original random operators 
 by replacing each random variable $n_R$ by  the operators $\widehat{\bf M}_R$.
It is an operator in the augmented space 
$\mathbf{\Xi}$ = ${\cal H} \otimes \mathbf{\Psi}$. The theorem maps 
a disordered operator described in a Hilbert space ${\cal H}$ onto an ordered one 
in an enlarged space $\mathbf{\Xi}$, where this space  is constructed as the outer product of the
space ${\cal H}$ and configuration space $\mathbf{\Psi}$ of the random variables.
 The  configuration space $\mathbf{\Psi}$ is of rank 2$^{N}$ if there are $N$ sites 
 in the system. Another way of looking at the augmented space operators is to note
that they are {\sl collection} of all possible operators for all possible
configurations of the system.

The augmented space operators may be written as \cite{alam}:

\begin{widetext}
\begin{eqnarray}
\widehat{\mathbf m} &=& \ll \mathbf{m}\gg \hat{\mathbf{I}} 
+ (\mathbf{m}_A-\mathbf{m}_B)\sum_R {\cal P}_R\otimes \left[(y-x)\ {\cal P}^\downarrow_R 
\pls \sqrt{xy} \ {\cal T}^{\uparrow\downarrow}_{R}
\right],\nonumber\\
\widehat{\mathbf \Phi}_{off} & =& \sum_{R}\sum_{R\ne R'} {\cal T}_{RR'}\otimes \left[\rule{0mm}{4mm}
\ll\mathbf{\Phi}_{RR'}\gg {\cal I} +  
\mathbf{\Phi}^{(1)}_{RR'}\left\{\rule{0mm}{3mm} (y-x)\left({\cal P}_R^\downarrow + {\cal P}_{R'}^\downarrow\right)+
\sqrt{xy}\left({\cal T}^{\uparrow\downarrow}_R+ {\cal T}^{\uparrow\downarrow}_{R'}\right)\right\} +\right. \nonumber\\
 & &  \left.  \mathbf{\Phi}^{(2)}_{RR'}\left\{\rule{0mm}{3mm} \sqrt{xy}(y-x) \left({\cal P}^\downarrow_R\otimes{\cal T}^{\uparrow\downarrow}_{R'} + {\cal P}^\downarrow_{R'}\otimes{\cal T}^{\uparrow\downarrow}_{R}\right)
+ (y-x)^2 {\cal P}^\downarrow_{R}\otimes{\cal P}^\downarrow_{R'} 
+ xy {\cal T}^{\uparrow\downarrow}_{R}\otimes{\cal T}^{\uparrow\downarrow}_{R'}\right\}\rule{0mm}{4mm}\right],
\nonumber\\
\widehat{\mathbf \Phi}_{dia} & =&  \sum_{R} {\cal P}_{R}\otimes \left[- \ll\mathbf{\Phi}_{RR}\gg {\cal I} -\sum_{R'\ne R}\mathbf{\Phi}^{(1)}_{RR'}\left(\rule{0mm}{3mm}
(y-x){\cal P}_R^\downarrow + \sqrt{xy} {\cal T}_R^{\uparrow\downarrow}\right) 
 -\sum_{R'\ne R} \mathbf{\Phi}^{(1)}_{RR'}\left(\rule{0mm}{3mm} (y-x){\cal P}_{R'}^\downarrow+
\sqrt{xy}{\cal T}^{\uparrow\downarrow}_{R'}\right)\right. \nonumber\\
 & & \left. -\sum_{R'\ne R} \mathbf{\Phi}^{(2)}_{RR'}\left\{\rule{0mm}{3mm} \sqrt{xy}(y-x) \left({\cal P}^\downarrow_R\otimes{\cal T}^{\uparrow\downarrow}_{R'} + {\cal P}^\downarrow_{R'}\otimes{\cal T}^{\uparrow\downarrow}_{R}\right)
+ (y-x)^2 {\cal P}^\downarrow_{R}\otimes{\cal P}^\downarrow_{R'} 
+ xy {\cal T}^{\uparrow\downarrow}_{R}\otimes{\cal T}^{\uparrow\downarrow}_{R'}\right\}\rule{0mm}{4mm}\right],\nonumber\\
\widehat{\mathbf \Phi}&=&\widehat{\mathbf \Phi}_{dia}+\widehat{\mathbf \Phi}_{off},
\label{eq24}
\end{eqnarray}
\end{widetext}
where, 
\begin{eqnarray*}
\mathbf{\Phi}^{(1)}_{RR'} & = & x\mathbf{\Phi}^{AA}_{RR'}-y\mathbf{\Phi}^{BB}_{RR'}+(y-x)\mathbf{\Phi}^{AB}_{RR'},\\
\mathbf{\Phi}^{(2)}_{RR'} & = & \mathbf{\Phi}^{AA}_{RR'}+\mathbf{\Phi}^{BB}_{RR'}-2\mathbf{\Phi}^{AB}_{RR'}.
\end{eqnarray*}

\noindent The sum rule which connects the diagonal and off-diagonal parts of the force-constant matrices has been incorporated into the formulation.
 
\section{The Itinerant CPA and Augmented space recursion}

The augmented space theorem described in the previous section is an {\sl exact} statement. It is a clever
book keeping technique to include the effects of disorder fluctuations in the model of phonons in 
our random alloy. However, it is {\sl not} an algorithm for the approximate calculations of  spectral and
other physical properties of phonons in disordered alloys.  For that we have to turn to either mean-field
approximations like the CPA and ICPA or alternatively to the ASR. The  coherent potential like mean-field
approximations begin with a {\sl partition} of the augmented space into a part which is spanned by the {\sl reference} or {\sl null
cardinality} state $\vert \{\emptyset\}\rangle$ which we shall call the {\sl average} configuration state
and the remaining part $\mathbf{\Psi}- \vert\{\emptyset\}\rangle\langle\{\emptyset\}\vert$ spanned by
{\sl fluctuation} states : $\{\vert \{{\cal C}\}\rangle\}$.  With this partition, any operator can be written
in a block representation :

\[\mathbf{A} \ =\ \left( \begin{array}{cc}
         \mathbf{A}_1 & \mathbf{A}' \\
         \mathbf{A}'^\dagger & \mathbf{A}_2
	\end{array}\right). \]

The partition or downfolding theorem then allows us to invert this operator in the subspace spanned by
the {\sl average} configurations alone. By the augmented space theorem this is the configuration average.
If we define the operator  {\bf K} as ({\bf m}$\omega^2-{\mathbf \Phi}$), then using the above partition~: 
{\bf K}$_1 = (\ll{\mathbf m}\gg\omega^2-\ll{\mathbf \Phi}\gg)$ . The downfolding theorem and augmented space theorem together give us~:

\begin{eqnarray}
\ll {\mathbf G}(\omega^2)\gg &=& (\mathbf{K}_1-\mathbf{K}'^\dagger\ \mathbf{F}\ \mathbf{K}')^{-P_1},\nonumber\\
 &= & (\mathbf{G}_{VCA}^{-1}(\omega^2)-{\mathbf \Sigma}(\omega^2))^{-P_1},
  \nonumber\\
\mathbf{F} & = &  \mathbf{K}_2^{-P_2} \quad\mbox{is the itinerator}\\
\mathbf{\Sigma} & = & \mathbf{K}'^\dagger\ \mathbf{F}\ \mathbf{K}'\quad
\mbox{is the self-energy}
\end{eqnarray}

\noindent Here {\bf A}$^{-P_1}$ and {\bf A}$^{-P_2}$ refer to the inverses of the operator {\bf A} in the subspaces
labelled by 1 and 2. This is exactly the partitioning idea introduced by Srivastava {\em {et al}} \cite{skm}. Ghosh
{\em {et al}} \cite{glc} next confined themselves to {\sl single fluctuation} states of the type $\vert\{R\}\rangle$
and went ahead to self-consistently evaluate the self-energy in this approximation.
Adopting their notation $\langle \{R\}\vert {\mathbf A}\vert \{R'\}\rangle$ = $ A^{(R){(R')}}$, they
used translational symmetry in augmented space \cite{trans} and approximated the 
self-energy and {\sl itinerator} {\bf F} within the single fluctuation states :

\begin{eqnarray}
\mathbf{\Sigma} & = & \sum_{RR'}{\mathbf{K}'^\dagger}^{(R)}\ \mathbf{F}^{(R)(R')}\ \mathbf{K}'^{(R')},\\
\mathbf{F}^{(R)(R')} & =&  \mathbf{G}^{(R)}\left[\delta_{RR'} + \sum_{R^{\prime\prime}} \mathbf{V}^{(R)(R^{\prime\prime})}\ F^{(R^{\prime\prime})(R')}\right].
\end{eqnarray}

In going from the  equation (5) to (7) all contributions to the self-energy
of configuration states with more than one fluctuations in more than one site
have been neglected. Similarly in going from equation (6) to (8), matrix elements of the itinerator {\bf F} between configuration states with more than one fluctuation present at a time, which corresponds to coherent scattering from more than one site have been neglected and such states do not contribute to {\bf F} and hence to the self-energy $\mathbf{\Sigma}$ within this approximation.
 The second equation is a Dyson equation within the subspace spanned by  only single fluctuation states.
Self-consistency is achieved through :

\begin{eqnarray*}
\mathbf{G}^{(R)} &=& \left(\mathbf{G}^{-1}_{VCA}-\mathbf{\Sigma}^{(R)}\right)^{-1},\\
\mathbf{\Sigma}^{(R)}& =& \sum_{R'R^{\prime\prime}\ne R} \mathbf{K}'^{(R')}\ \mathbf{F}^{(R')(R^{\prime\prime})}\ \mathbf{K}'^{(R^{\prime\prime})}.
\end{eqnarray*}

The above argument shows that unlike the usual CPA where only a single fluctuation at a site is considered, multiple fluctuations coming from multiple-scattering is present in the itinerator {\bf F} and therefore contribute to the self-energy $\mathbf{\Sigma}$. However, the approximation described above means that correlated fluctuations
between more than one site are present in neither the itinerator nor the self-energy. This is the main
approximation involved in the ICPA.

We note that the ICPA is a self-consistent mean-field approximation for the self-energy which relates
the configuration averaged Green function to the virtual crystal one. It is an approximation which maintains
both the translational symmetry of the configuration average and its herglotz analytic properties. The ASR is an alternative technique for doing the same thing, namely obtaining an approximation to the self-energy maintaining the necessary properties of the exact case.

The recursion method addresses inversions of infinite matrices \cite{vol35}. 
The average Green function in the augmented space formalism can be written as \cite{alam} :
\[\ll \mathbf{G}\gg = \{1\vert (\omega'^2 \hat{\mathbf{I}} - \widetilde{\mathbf{K}})^{-1}\vert 1\}\],
Here 
\begin{eqnarray*}
\omega' & = & \mu^{1/2} \omega, \\
 \widehat{\mathbf K}& = &
\left(\widehat{\mathbf{m}^{-1/2}}\mu^{1/2}\right)\widehat{\mathbf{\Phi}} \left(\widehat{\mathbf{m}^{-1/2}}\mu^{1/2}\right),\\ 
\vert 1\} & =& \left(\mu_1^{-1/2}\mu^{1/2}\right)\vert \{\emptyset\}\rangle + 
\left(\mu_2^{-1/2} \mu^{1/2}\right)\vert \{R\}\rangle, 
\end{eqnarray*}
Where $\mu^{1/2} = \ll m^{-1}\gg^{-1/2}$, $\mu_1^{-1/2} = \ll m^{-1/2}\gg$ and
$\mu_2^{-1/2} = \sqrt{xy}((m_A^{-1/2}-m_B^{-1/2})$.

Once a sparse representation
of an operator in a Hilbert space, ${\bf\widehat K}$, is known in a countable basis, the recursion method
obtains an alternative basis in which the operator becomes tridiagonal. 
This basis and the 
representations of the operator in it are found recursively through a three-term recurrence relation~:
\begin{equation}
|u_{n+1}\} = {\bf\widehat K } |u_n\} - \alpha_n(\k) |u_n\} - \beta_n^2(\k) |u_{n-1}\}.
\end{equation}
\vskip 0.2cm
with the initial choice $|u_1\}=|\k\otimes \{1\}\rangle$  or and $\beta_1^2=1$. The recursion 
coefficients $\alpha_n$ and $\beta_n$ are real and are obtained by imposing the ortho-normalizability 
condition of the new basis set as~:
\begin{eqnarray*}
&&\alpha_n(\k) = \frac{\{u_{n}|{\bf\widehat K}|u_{n}\}}{\{u_{n}|u_{n}\}} \phantom{x} ; \phantom{xx} \beta_{n+1}^2(\k) = \frac{\{u_{n}|{\bf\widehat K }|u_{n+1}\}}{\{u_{n}|u_{n}\}} \phantom{x} \\
&& \mbox{and also} \phantom{xx} \{u_{m}|u_{n}\} = 0 \mbox{  for } m\not= n, n\pm 1.
\end{eqnarray*}
Now, we use the augmented space theorem and repeated applications of the downfolding theorem on the
tri-diagonal representation gives :

\begin{widetext}
\begin{equation}
 \ll \mathbf{G}(\k,\omega'^2) \gg \ = \ \frac{1}
        {\displaystyle \omega'^2-\alpha_{1}(\k)-\frac{\beta^2_{2}(\k)}
        {\displaystyle \omega'^2-\alpha_{2}(\k)-\frac{\beta^2_{3}(\k)}
        {\displaystyle \frac{\ddots}
        {\displaystyle \omega'^2-\alpha_{N}(\k)-\mathbf \Gamma(\k,\omega'^2)}}}}
\ =\ \frac{1}{\displaystyle\omega'^2-\alpha_{1}(\k)-\Sigma '(\k,\omega'^2)}.
\end{equation}
\end{widetext}

From the definition of the self-energy given earlier, it has been argued by us in an
earlier communication that the disorder scattering induced lifetimes come entirely 
from the imaginary part of $\Sigma '(\k,\omega'^2)$. 
Here $\mathbf \Gamma(\k,\omega'^2)$ is the asymptotic part of the continued fraction.  
The {\it approximation} involved has to do with the termination of this continued fraction. The coefficients 
are calculated exactly up to a finite number of steps $\{\alpha_n,\beta_n\}$ for $n < N$ and the asymptotic
part of the continued fraction is obtained from the initial set of coefficients using the idea of Beer and Pettifor terminator \cite{beer-pet}. With this terminator, the approximate Green function maintains the herglotz properties of the exact result. Haydock and co-workers \cite{hhk} have carried out extensive studies of 
the errors involved and precise estimates are available in the literature. Haydock \cite{hhk} has shown
that if we carry out recursion exactly up to $N$ steps, the resulting continued fraction maintains the first 
$2N$ moments of the exact result.

Both the ICPA and the ASR involve approximations of the self-energy. We have already discussed that in the ICPA contributions of configurations involving correlated fluctuations in more than one site to the self-energy are ignored in
the present case but it is capable of incorporating them. If we use the form of Eqn. (4) in the
recursion equations (9), it is immediately obvious that in the ASR, contributions of such
correlated fluctuation states to the self-energy are present. We had earlier shown that
such contributions occur first at $\beta_4^2$ for diagonal disorder and in $\alpha_2$ in case of off-diagonal disorder. Ignoring such contributions will make all moments greater than or equal to eight to be non-exact for digonal and three for off-diagonal disorder.
The ICPA achieves accuracy through self-consistency in the subspace of single fluctuations in it's present version , while the ASR achieves accuracy by increasing the number of recursions in the
{\sl full} augmented space  and estimating
the terminator to mimic the asymptotic part of the continued fraction as closely as possible. The two are very different algorithms. Both can take care of off-diagonal disorder and short-ranged order, but in those situations where clustering, either chemical or statistical, is important, the ASR, which takes into account correlated scattering from clusters, should be preferable over the
single-fluctuation only version of the ICPA.

\section{First principles calculations of force constants in alloys}

As is clear from the above discussions that the crucial component in both
the ASR and the ICPA is the alloy force constants. Due to the random
chemical environment around each atom in a substitutionally disordered
alloy, the force constants corresponding to A-A,B-B and A-B pairs in
a A$_{x}$B$_{1-x}$ alloy are different and in no way resembles the force
constants in a completely ordered environment. In order to have significant
accuracy in calculated phonon properties one should, therefore, have 
accurate information on force constants corresponding to various pairs of
chemical species. The only trustworthy source of force constant data is
the first-principles calculation. To this end, we have employed 
first-principles Density functional perturbation theory (DFPT) to obtain
force constants. The details about the DFPT and our approach to use it
to extract random alloy force constants is discussed below.

\subsection{Density functional perturbation theory}

Density functional perturbation theory (DFPT) \cite{dfpt} 
is a density functional
theory (DFT) based linear response method to obtain the electronic
and lattice dynamical properties in condensed matter systems. The 
dynamical matrix which provides information on lattice dynamics of the
system can be obtained from the ground-state electron charge density and
it's linear response to a distortion of the nuclear geometry \cite{johnson}. 
In DFPT,
this linear response is obtained within the framework of DFT. One of
the greatest advantages of DFPT-as compared to other nonperturbative
methods for calculating lattice dynamical properties of crystalline solids
(such as the frozen-phonon or molecular dynamics spectral analysis methods)
-is that within DFPT the responses to perturbations of different wavelengths
are decoupled. This feature allows one to calculate phonon frequencies
at arbitrary wavevectors avoiding the use of supercells and with a
workload that is independent of the phonon wavelength.

\subsection{Random alloy force constants from DFPT}

Since there is no first-principles theory for lattice dynamics in random
alloys available we took recourse to calculate force constants for
ordered structures which can suitably mimic the random alloy using DFPT.
However, for a proper representation for the random alloy, one needs
to work with a large supercell which prohibits the use of DFPT from a
practical point of view. The other approach would have been to construct
a set of ordered structures having the same composition of the alloy under
investigation, run first-principles calculation on each of them and 
average the data appropriately. As a first approximation to this approach,
here we have done DFPT calculations on a single ordered structure and
used the resultant force constants as approximate random alloy force 
constants as inputs to the ICPA and the ASR. The alloy chosen is FePd.
The reason for
choosing the FePd system is twofold: first, the ICPA and the ASR
were applied only for the NiPt and NiPd alloys where the
constituents of the alloys have face-centred cubic structure in their
elemental phases. In case of FePd, although the alloy in the disordered
phase is face centred cubic but Fe is body centred cubic in it's elemental
phase. It was therefore interesting to test the suitability of both
the approximations in case of such a system where one of the constitutents
forming the alloy has a different structure than the alloy itself in it's
elemental phase. second, inelastic neutron scattering data was available
for Fe$_{50}$Pd$_{50}$ \cite{expt}. It would therefore have been possible
to compare the ICPA and the ASR results with the experimental data
directly enabling the understanding of the nature of interactions between
various pairs of species in the random phase. Since the 
Fe$_{50}$Pd$_{50}$ forms a face centred cubic (fcc) solid solution,
we have chosen 
the prototype tetragonal L1$_{0}$ structure with c/a ratio equal to unity
to be used for first-principles calculations.

\subsection{Details of first-principles calculations}
We use DFPT within the local-density-approximation (LDA) to compute
the force constants for the FePd equiatomic
composition single ordered structure mentioned above.
The experimental lattice constant $a=7.24$ a.u. is used in the 
calculations. We employ a plane-wave pseudopotential implementation
of the DFPT with Perdew-Zunger parametrization of the LDA \cite{pz} as done
in the Quantum-Espresso package \cite{espresso}. 
Ultrasoft pseudopotentials \cite{dhv} with
non-linear core correction \cite{nlcc} 
are used for Fe and Pd. The kinetic energy
cut-off is taken to be 35 Ry. The Brillouin-zone integrations are carried
out with Methfessel-Paxton smearing \cite{mp} 
using a 10$\times$ 7$\times$ 7$\times$
{\bf k}-point mesh, which corresponds to 120 {\bf k}-points in the
irreducible wedge. The value of the smearing parameter is 0.1 Ry. These
parameters are found to yield phonon frequencies converged to within
5 cm$^{-1}$.

Once adequate convergence is achieved for the electronic structure, the
phonon force constants are obtained using the linear response. Within
DFPT, the force constants are conveniently computed in reciprocal space
on a finite ${\bf q}$-point grid and Fourier transformation is employed
to obtain the real-space force constants. The number of unique real-space
force constants and their accuracy depend upon the density of the 
${\bf q}$-point grids: the closer the ${\bf q}$-points are spaced, the 
accurate the force constants are. In this work, the dynamical matrix is
computed on a 6$\times$ 6$\times$ 4 ${\bf q}$-point mesh \cite{fc} commensurate
with the ${\bf k}$-point mesh.

\section{Results and Discussions}

In Table I we report the nearest-neighbor force constants for the artificial
ordered structure A-B obtained from first-principles as described above.
\begin{table}[t]
\caption{Real-space nearest neighbor force constants for Fe$_{50}$Pd$_{50}$
obtained by DFPT calculations on the artificial ordered structure. The units
are dyn cm$^{-1}$.}
\begin{center}
\begin{tabular}{lcr}
\hspace{0.67in} & \hspace{0.67in} & \hspace{0.67in} $\,$\\
Pair & Force constant & Direction  \\
\hline
\hline
Fe-Fe & -9458 & 1$xx$\\
Fe-Pd & -9458 & 1$xx$\\
Pd-Pd & -28974 & 1$xx$\\
\hline
Fe-Fe & -6005 & 1$xy$\\
Fe-Pd & -10755 & 1$xy$\\
Pd-Pd & -30372 & 1$xy$\\
\hline
Fe-Fe & 1800 & 1$zz$\\
Fe-Pd & -114 & 1$zz$\\
Pd-Pd & 3555 & 1$zz$\\
\hline
\hline
\end{tabular}
\end{center}
\end{table}
Subsequently, we use these force constants as inputs to the ICPA and the 
ASR calculations.
\begin{figure}[b]
\includegraphics[width=8.5cm,height=11cm]{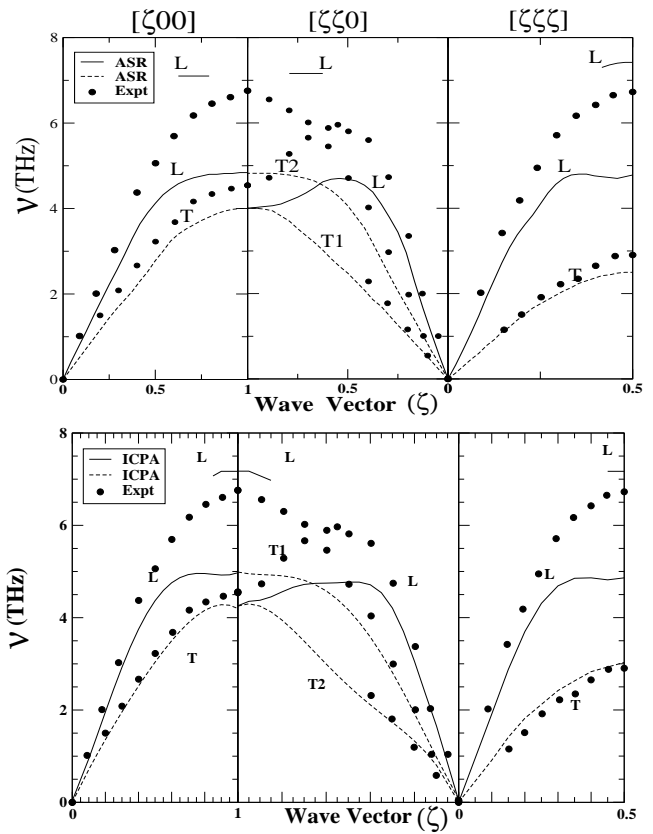}
\caption
{Dispersion Curves (\ frequency $\nu$ vs. reduced wave vector $\zeta$\ ) for Fe$_{50}$Pd$_{50}$ alloy. The upper panel correponds to the ASR results however the lower panel to the ICPA results. The filled circles are the experimental data \cite{expt}.  The force constants used are given in Table I. }
\label{fig1}
\end{figure}
Figure (\ref{fig1}) shows the corresponding dispersion curves. The results clearly
show that the force constants for the artificial ordered structure are not
adequate to describe the lattice dynamics for the disordered Fe$_{50}$Pd$_{50}$
system. In both the ICPA and the ASR, the high frequency phonons are 
poorly represented for all three symmetry directions. On top of that, the
high frequency branches suffer a split for large $q$ values, a feature
not observed in the experiments. All these features point to the fact that
the force constants used in the ICPA and the ASR calculations completely
fail to capture the complexities of the force-constant disorder in a random
environment. This is quite understandable as we have used a crude approximation
for the force constants in random environment. One single ordered structure
, in no way, can mimic the randomness in the environment around a given
chemical species. The fact that the force constants obtained on this
artificial structure are responsible for the disagreement with the 
experiment is corroborated by the coherent structure factors, particularly
at high $q$ values as demonstrated in Figure (\ref{fig2}) where coherent structure
factors for certain high $q$ values, obtained by the ICPA, are displayed.
The curves clearly show that the spurious high frequency peak is due to
the Pd-Pd pairs and to a smaller extent due to Ni-Pd pairs. As is seen from
Table-I, the Fe-Fe and Pd-Pd force constants differ by about 70 $\%$, thereby
representing a situation of very strong disorder as is seen in the case of
Ni$_{50}$Pt$_{50}$ \cite{glc}. The splitting of the high frequency branch
is a manifestation of this strong force-constant disorder, albeit wrong 
in the present case. 
\begin{figure}
\includegraphics[width=8.5cm,height=8cm]{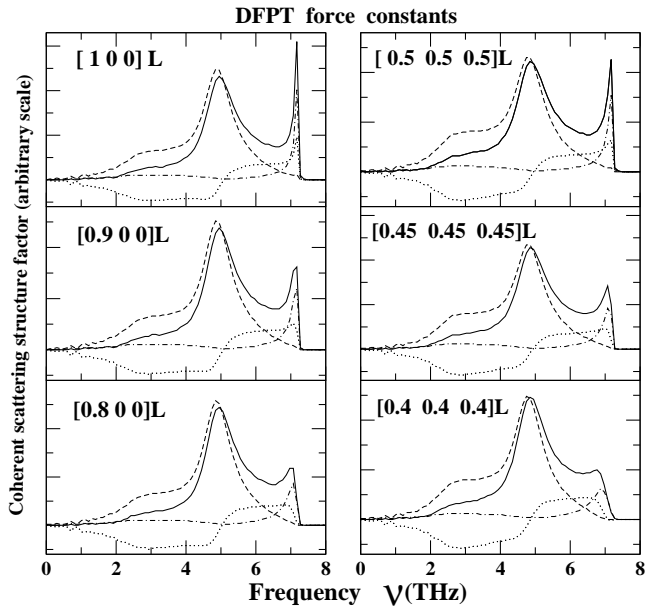}
\caption
{Partial and total structure factors calculated in the ICPA for various $\zeta$ values along the [$\zeta, 0, 0$] and [$\zeta, \zeta, \zeta$] directions in Fe$_{50}$Pd$_{50}$ alloy. The type of mode is labelled along a particular symmetry direction. The force constants used are given in Table I. The details are given in the text. {\bf  Description of various curves}}
\label{fig2}
\end{figure}
In the pursuit of the correct set of force constants for the system so that
the suitability of the ICPA and the ASR can be properly tested, we then 
used the force constant data as reported in the experiment \cite{expt}. 
However, the force constant data reported in the experiment was obtained
by fitting the frequencies to a Born-Von-Karman force constant model. 
Also, the frequencies were obtained from neutron-scattering data on 
ordered L10 structure at 860K, very close to the order-disorder transition
temperature 950K. The reason behind using the experimental force constants
obtained from an ordered L10 structure for disordered calculations were
twofold: first, the L10 force constants should be a better approximation
for random alloy force constants than the artificial cubic structure ones 
because the L10 structure allows structural relaxation and therefore a variation
in the bond distances between different chemical species pairs although
in a restricted way. Nevertheless, this restricted degree of relaxation
could be crucial in capturing the nature of forces between various chemical
species as has been seen in case of NiPt alloys \cite{sg}. Second, since
the L10 data was taken at 860K and the disordered fcc data was taken at 
1020K, both of them lie very close to the order-disorder transition 
temperature. At a first-order order-disorder transformation at finite
temperatures, the ordered phase is only partially ordered and the
disordered phase is in equilibrium with, has short-range order. 
\begin{figure}[t]
\includegraphics[width=8.5cm,height=11cm]{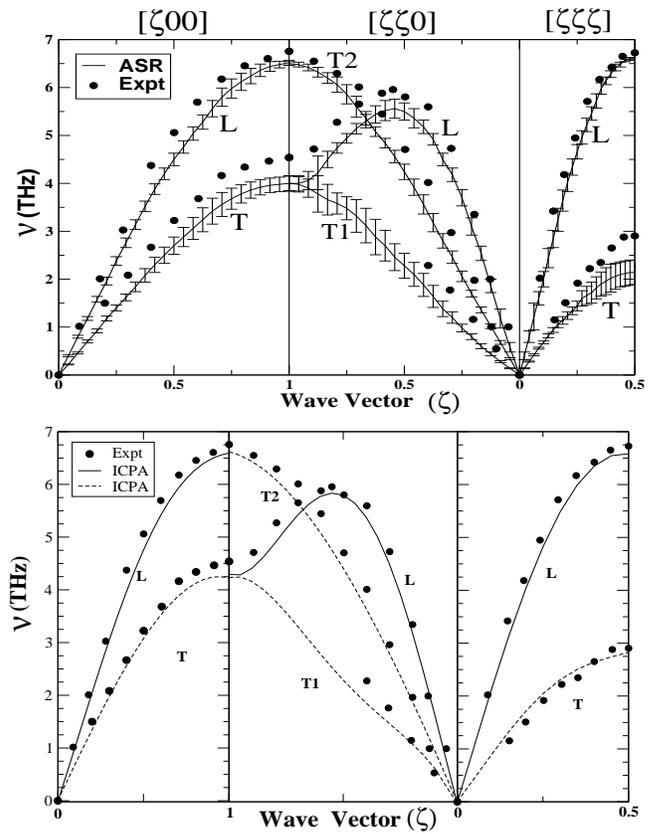}
\caption
{Dispersion Curves (\ frequency $\nu$ vs. reduced wave vector $\zeta$\ ) for Fe$_{50}$Pd$_{50}$ alloy. The upper panel correponds to the ASR results however the lower panel to the ICPA results. The filled circles are the experimental data \cite{expt}. The error bar in the ASR result basically represents the full widths at half maxima (FWHM) at various $\zeta$ values.  The force constants used are given in Table II. }
\label{fig3}
\end{figure}
Examination of the correlation functions has shown that ordered and disordered states 
rather show similar atomic arrangement in the vicinity of the order-
disorder transformation. It is therefore expected that in the present case,
the L10 force constants at 860K would not change significantly in the
disordered phase at 1020K. These intuitive arguments are well supported by
the dispersion curves presented in Figure (\ref{fig3}). Both the ICPA and the ASR
results agree reasonably well with the experimental data. The spurious
splitting obtained earlier disappears. This disappearance can be understood
better if we look at the force constants used for this calculation. Table-II
lists the experimental force constants used as inputs for the ICPA and the
ASR calculations.
\begin{table}[t]
\caption{Real-space nearest neighbor force constants for Fe$_{50}$Pd$_{50}$
obtained from experimental data \cite{expt} on L10 structure at 860K. The units
are dyn cm$^{-1}$.}
\begin{center}
\begin{tabular}{lcr}
\hspace{0.67in} & \hspace{0.67in} & \hspace{0.67in} $\,$\\
Pair & Force constant & Direction  \\
\hline
\hline
Fe-Fe & -5650 & 1$xx$\\
Fe-Pd & -14050 & 1$xx$\\
Pd-Pd & -19450 & 1$xx$\\
\hline
Fe-Fe & -9750 & 1$xy$\\
Fe-Pd & -16550 & 1$xy$\\
Pd-Pd & -22350 & 1$xy$\\
\hline
Fe-Fe & 4100 & 1$zz$\\
Fe-Pd & 2500 & 1$zz$\\
Pd-Pd & 2900 & 1$zz$\\
\hline
\hline
\end{tabular}
\end{center}
\end{table}

\begin{figure}[b]
\includegraphics[width=8.5cm,height=9cm]{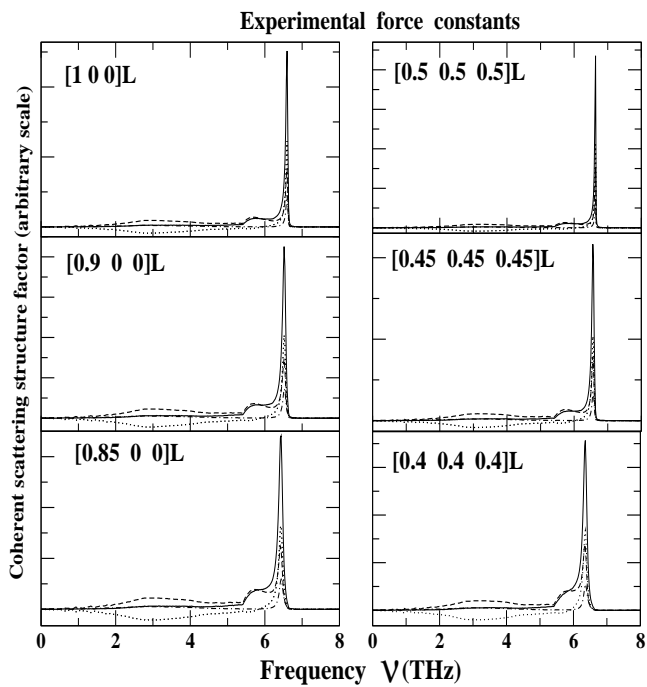}
\caption
{Partial and total structure factors calculated in the ICPA for various $\zeta$ values along the [$\zeta, 0, 0$] and [$\zeta, \zeta, \zeta$] directions in Fe$_{50}$Pd$_{50}$ alloy. The type of mode is labelled along a particular symmetry direction. The force constants used are given in Table II. Other details are given in the text. {\bf Description of various curves}}
\label{fig4}
\end{figure}
In comparison with DFPT values, the Pd-Pd force constants are a lot 
softer and the Fe-Pd force constants harden. The fact that the force 
constants and their behavior is indeed the deciding factor is again
exemplified by the coherent structure factors for the selected 
high $q$ vectors as shown in Figure (\ref{fig4}). The figures show that the
single high frequency peak is now mostly because of the Fe-Fe and 
Fe-Pd contributions, rather than the Pd-Pd contribution. This points to
the fact that the Pd-Pd contribution was overestimated and the
Fe-Fe and the Fe-Pd contributions were grossly underestimated by the
DFPT calculations on the artificial cubic structure because of the lack
of relaxation in such structure. This in turn can be understood by
looking at the bond distances between various pairs of species. In
the cubic structure, the Fe-Fe, Pd-Pd and the Fe-Pd distances were 
same and in the present case was taken to be 5.12 a.u. The L10 structure
at 860 K, on the other hand, had Fe-Fe and Pd-Pd distances to be 5.25 a.u.
and the Fe-Pd distances to be 5.08 a.u. Thus, the Pd atoms, in the artificial
cubic structures, were made to vibrate in a smaller volume and 
because of the smaller distance between two like atoms, the Pd-Pd force 
constants became harder.
\begin{figure}[b]
\includegraphics[width=8.2cm,height=10cm]{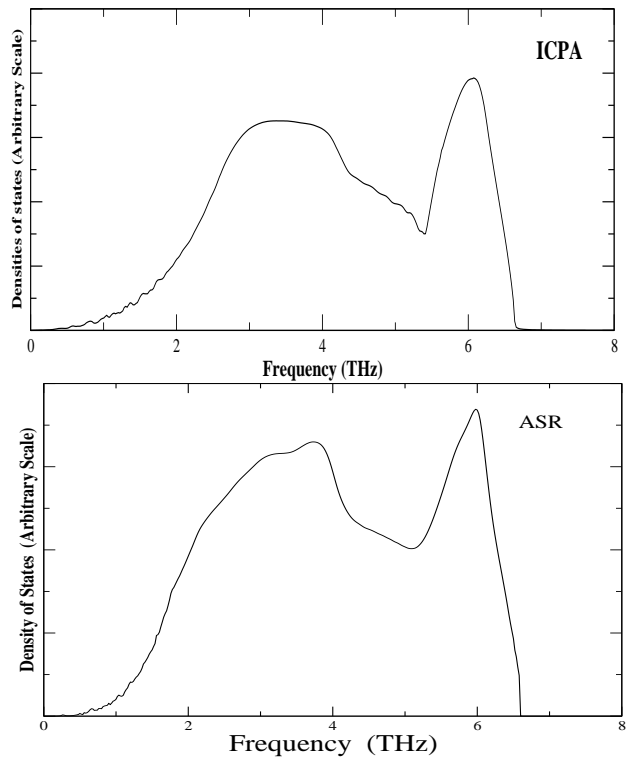}
\caption
{Phonon density of states for Fe$_{50}$Pd$_{50}$ alloy. The upper and lower panel shows ICPA and ASR results respectively. The force constants used are that of experimental paper \cite{expt} given in Table II.}
\label{fig5}
\end{figure}
\begin{figure}[h]
\includegraphics[width=8.5cm,height=13.5cm]{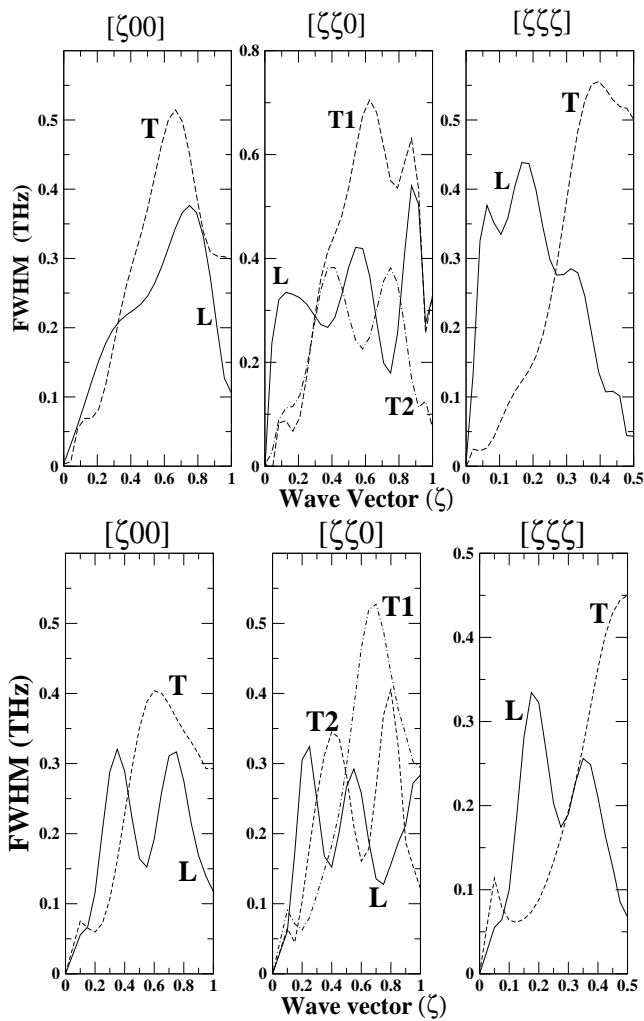}
\caption{Disorder-induced FWHM's vs wave vector ($\zeta$) for Fe$_{50}$Pd$_{50}$ alloy. The upper and lower panels show ASR and ICPA results respectively. The force constants used are given in Table II. }
\label{fig6}
\end{figure}
Figure (\ref{fig5}) compares the ICPA and the ASR results for the phonon densities of
states. Both the approximations produce identical features. The peaks and
the band edges have quantitative agreement among themselves and with
the experimental results \cite{expt}. 

Figure (\ref{fig6}) displays the Full width at half maxima (FWHM) data associated with 
the finite lifetimes of phonons due to disordered scattering. The upper three panels show the widths as a function of wave vector ($\zeta$) along the three symmetry directions extracted from the ASR method, however the lower three panels show the ICPA results. The FWHM is a more sensitive test for the underlying approximation than the dispersion curves. The ICPA and the ASR have reasonable agreement regarding FWHM's although quantitative agreement is not there because of the different nature of the two approximations. Unfortunately, the experimental group didn't perform a phonon life-time measurement so that the FWHM's calculated theoretically could be compared with the experiments.
\begin{widetext}
\begin{figure*}
\centering
\includegraphics[width=13cm,height=14cm]{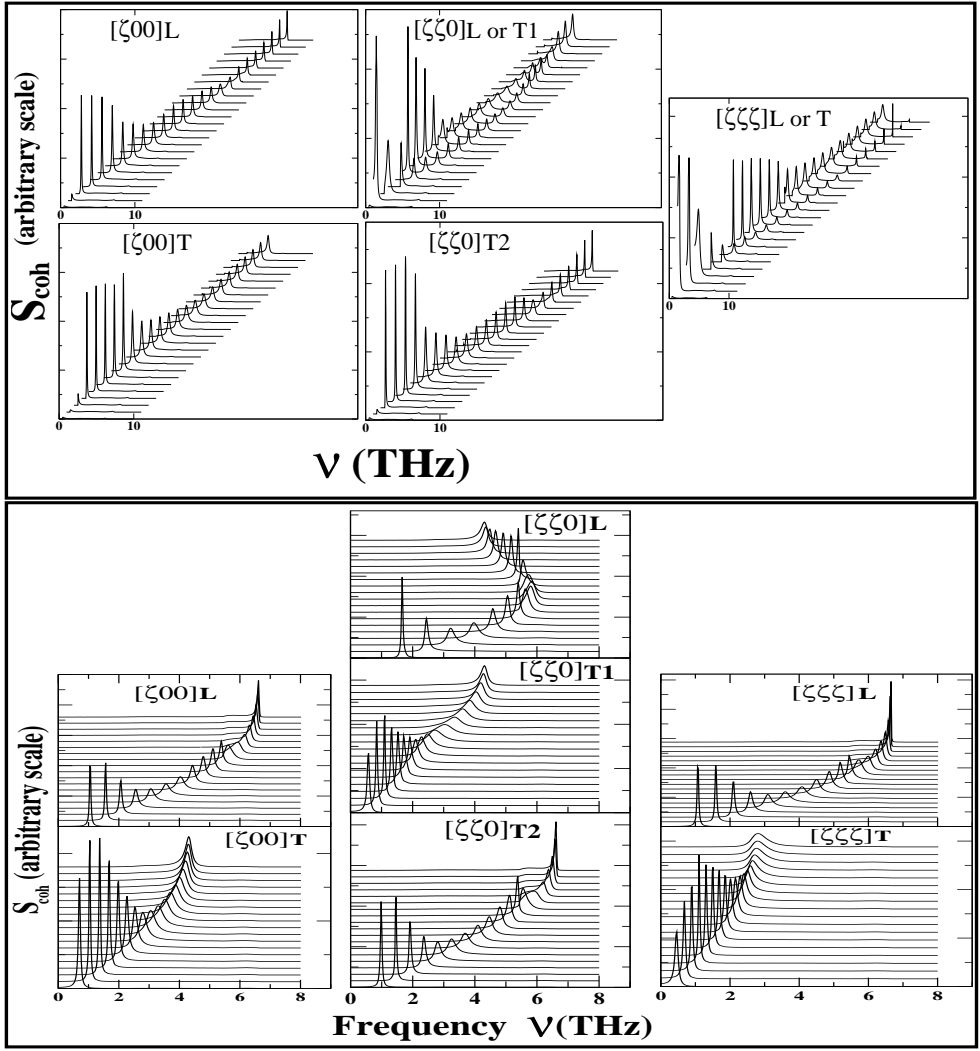}
\caption
{Total coherent structure factors in different directions with different branches for Fe$_{50}$Pd$_{50}$ alloy. The upper and lower box shows ASR and ICPA results respectively. In each of the different directions and branches, the various curves indicate the total structure factors for various $\zeta$ values starting from the lowest value to the edge of the Brillouin zone. In the ASR result, the T1 and T2 modes are degenerate along [$\zeta 0 0$] direction, L and T1 modes are degenerate along the [$\zeta \zeta 0$] direction, however all the three modes are degenerate along the [$\zeta \zeta \zeta$] directions. In the ICPA result, non of the modes are degenerate. Such differences in the two methods are hidden inside the structure of two different calculations. In both the boxes, the different curves for different $\zeta$ values are shifted along the x-axis in order to facilitate vision.  }
\label{fig7}
\end{figure*}
\end{widetext}
In Fig. (\ref{fig7}), we have shown the total coherent structure factors at various $\zeta$ values along the three symmetry directions with various modes of vibrations. The upper box shows ASR result however the lower ICPA result. In both the cases, the different curves for different $\zeta$ values are shifted along the x-axis in order to facilitate vision. One can easily notice the difference in the nature of curves arising out of two different methodologies. In the ASR result, we have found three classes of degenerate modes, these are (1)\ T1 and T2 modes along [$\zeta 0 0$] directions\ (2)\ L and T1 modes along [$\zeta \zeta 0$] directions and\ (3)\ all the three modes along [$\zeta \zeta \zeta$] directions. However in the ICPA case, all the modes along different directions are non-degenerate. The difference in the two kinds of results are due to the different structure and  way of calculations in the two methodologies. However the ultimate dispersion curves and FWHM's came out from two different methodologies are reasonably well compareable, which are the actual vibrational properties to look at in any disordered alloy.

\section{CONCLUSIONS}
This paper has continued the development of the Augmented space recursion \cite{alam} and the Itinerant coherent potential approximation (ICPA)\cite{glc} for studying the vibrational properties of disordered metallic alloys. A brief description of the two methods combined with a first-principles calculation (the so called DFPT) of the dynamical matrices has been reported. The power of these approaches has been illustrated by explicit calculations on the Fe$_{50}$Pd$_{50}$ alloy.

Both the theories are unique and systematic in the sense that they produce almost identical results for a particular system. Both the theories can explicitly take into account the fluctuations in masses, force constants and scattering lengths. We propose the methods as computationally fast and accurate techniques for the study of lattice dynamics of disordered alloys. A correct quantitative trend (compareable to the experimental results) of the phonon dispersion and the phonon density of states has been predicted by both the methodologies when the experimental force constants has been used in the calculation. Off-course there is a fairly obvious general comment to be made with regard to the self consistency of the procedure. This is precisely the reason that a first-principles estimate of the dynamical matrices on parent ordered alloys do not yield quantitatively accurate results (in comparison with the experiment) for the disordered alloy. We shall propose that we need to go beyond and estimate the dynamical matrices from a model of embedded atoms in a fully disordered back ground.

\section*{Acknowledgment}
One of us (AA) would like to acknowledge the CSIR for financial assistance in the form of a Fellowship Grant.


\begin{thebibliography}{99}
\bibitem{her} A complex function f(z) is called {\sl herglotz} if (i) all its
singularities lie on the real z-axis (ii) Im f(z) is negative everywhere in the
plane Im (z) $>$ 0 and is positive everywhere in Im (z) $<$ 0 (iii)  $\lim_{z=E\rightarrow\pm\infty} \Re e f(E) = 0,\ \mbox{E is real.}$
\end{thebibliography}
\end{document}